\documentclass[ 
  twocolumn, 
  floatfix,
  aps, 
  prx, 
  final,
  amsmath, amsfonts, amssymb
]{revtex4-2}

\usepackage{graphicx}
\usepackage{hyperref}
\usepackage[utf8]{inputenc}
\usepackage{physics}
\usepackage{xcolor}
\usepackage{lipsum}  

\newtheorem{theorem}{Theorem}

\begin{document}

\title{Epidemic reproduction numbers in spatial networks}

\author{Zahra Ghadiri}
\author{Jari Saram\"{a}ki}
\author{Takayuki Hiraoka}
\affiliation{Department of Computer Science, Aalto University, 00076 Espoo, Finland} 


\begin{abstract}

The basic and effective reproduction numbers are widely used metrics for characterizing the dynamics of infectious disease epidemics. However, the interpretation of these numbers is based on the assumption of homogeneous mixing and may not hold in real-world populations where the contact patterns deviate from that assumption. In this paper, we present a network-based framework to compare reproduction numbers in populations with and without spatial structure, while other parameters of the disease remain fixed. Using this framework, we show that in homogeneously mixed populations, in the absence of external interventions, the effective reproduction number decreases exponentially as the susceptible population declines. In contrast, in spatially structured populations, the basic reproduction number is smaller, and the effective reproduction number initially decreases faster but eventually converges to unity. We show that the reproduction number is determined by the level of competition between infectious nodes, which is governed by the network structure. Our results suggest that without knowledge of the network structure, reproduction numbers may not be informative for parameterizing the contagiousness of the disease or predicting the behavior of epidemic spreading. 
\end{abstract}

\maketitle

\section{Introduction}

The transmission of infectious diseases involves an intricate interplay of biological, social, and environmental factors, rendering epidemic dynamics inherently complex. This complexity poses challenges in understanding how outbreaks unfold, predicting their future course, and designing effective interventions. A widely adopted approach to address these challenges is to reduce epidemic behavior to a single representative quantity.
To characterize the epidemic dynamics at its initial stages, the basic reproduction number $R_0$ is commonly used~\cite{anderson_infectious_1991}. This number represents the expected number of infections caused by an initially infected individual in a fully susceptible population.

Beyond using $R_0$, monitoring real-time changes in disease transmission is crucial for assessing the effects of changes in contact patterns, intervention strategies, and the depletion of the susceptible pool. This can be achieved through the effective reproduction number, denoted by $R_\mathrm{eff}$~\cite{huisman_estimation_2022, flaxman_estimating_2020, pan_association_2020}, which represents the expected number of new infections caused by a single infected individual under the present circumstances~\cite{wallinga_different_2004}. The effective reproduction number is particularly valuable when a sizable fraction of the population has already been infected so that the assumptions behind $R_0$ no longer hold.  

These reproduction numbers are considered key variables in infectious disease epidemiology because they are associated with the threshold behavior of epidemics. Using branching process arguments, one can show that the disease invades a population and spreads widely when $R_0 > 1$, while it dies out rapidly otherwise. Similarly, an ongoing outbreak will continue to grow as long as $R_\mathrm{eff} > 1$, whereas the incidence will decline if $R_\mathrm{eff}$ falls below one. Thus, reproduction numbers provide a concise metric that informs public health authorities about the risks associated with infectious diseases and the effectiveness of control measures, such as vaccination~\cite{dhillon_getting_2020} and non-pharmaceutical interventions~\cite{flaxman_estimating_2020, jeyanathan_immunological_2020, thompson_control_2018}. 

Despite their widespread acceptance as a fundamental concept, however, limitations of reproduction numbers have also been recognized. One limitation is that  interpretations of reproduction numbers are deeply rooted in the assumption of a homogeneously mixed population, in which individuals interact randomly, and every infected individual has an equal chance of infecting every other. 
In reality, however, this assumption does not hold: individuals typically have contact with a small subset of all others in the population. A primary factor that determines this subset is spatial constraints, as proximity is required for disease transmission. These spatial constraints translate into a contact structure abundant in triangles, in which contact of $i$ with $j$ and $k$ implies contact between $j$ and $k$. 

This discrepancy in assumptions about the contact structure makes the interpretation of the reproduction numbers less straightforward than what the traditional approach offers~\cite{meyers_network_2005}. In the traditional approach, per-contact disease transmissibility and $R_0$ are directly proportional to one another.
However, with a more elaborate contact structure, this may not be the case. Furthermore,  mean-field models based on the homogeneous mixing assumption suggest that epidemics initially grow exponentially at a rate dictated by $R_0$, and subsequently $R_\mathrm{eff}$ exponentially decays to zero. Yet, in outbreaks of HIV/AIDS~\cite{colgate_risk_1989, szendroi_polynomial_2004},  Ebola~\cite{chowell_western_2014}, and COVID-19~\cite{thurner_networkbased_2020}, a polynomial spread has been observed. Under such a subexponential growth, the estimated effective reproduction number gradually approaches unity over time~\cite{chowell_characterizing_2016, kiskowski_modeling_2016}. However, these previous studies remain largely phenomenological and do not provide an explanation of the microscopic mechanisms behind this behavior of the reproduction number. 

In this paper, we focus on how the spatial constraints of the contact structure in the population affect the reproduction numbers while keeping per-contact transmissibility fixed. 
The spatial constraints can be incorporated by modelling the contact patterns as a network through which infection spreads~\cite{pastor-satorras_epidemic_2015, keeling_networks_2005, keeling_effects_1999}. We use analytical calculations and simulations to systematically compare reproduction numbers between stylized models of random and spatial networks. We demonstrate that the loopy contact structures induced by spatial constraints create competition between infectious individuals, reducing the basic reproduction number and driving the effective reproduction number towards unity.

This paper is organized as follows. In Section~\ref{sec: model}, we introduce the models for epidemic dynamics and contact networks, and define the reproduction number per generation $R_g$. In Section~\ref{sec: results}, we present our main results. We first derive the basic reproduction number $R_0$ for random geometric graphs and compare it to the corresponding $R_0$ for Erd\H{o}s-R\'{e}nyi graphs. We then analyze the behavior of $R_g$ as a function of $g$ for the two network models. Finally, we derive the reproduction number as a function of the force of infection for all networks interpolating across the spectrum of spatiality and degree heterogeneity. We discuss the implications of our findings and conclude in Section~\ref{sec: conc}.

\section{Models}\label{sec: model}
\subsection{Epidemic dynamics}
In this work, we model a population as an undirected contact network where the nodes represent individuals and the edges represent interactions between them through which the disease can be transmitted. We assume the contact network remains static throughout the epidemic. We focus on two categories of canonical compartmental epidemic models on contact networks: The susceptible-infectious-recovered (SIR) model and the susceptible-infectious (SI) model. In both models, the population is initially fully susceptible except for a randomly chosen infected node. Transmission occurs between an infected node and a susceptible node connected to it independently at a certain rate. After transmission, the susceptible node becomes infected and can infect its susceptible neighbors. In the SI model, an infected node will remain permanently infected, while in the SIR model, an infected node eventually transitions to the recovered state. A recovered node will no longer become infected or transmit the disease to others. Here, we assume Poisson processes with constant rates of infection and recovery~\cite{kenah_second_2007, karrer_message_2010}.

We adopt the epidemic percolation network (EPN) framework to implement the SI/SIR epidemic~\cite{kenah_second_2007}. An EPN is a directed network with edges denoting potential transmission from the tail node to the head node, provided that the tail is infected and the head is susceptible. This framework offers a way to map a stochastic epidemic process onto a static percolation model, thereby making the dynamics tractable.

More precisely, given an undirected contact network $G=(V, E)$, we encode stochastic transmission by an EPN as follows: 
For each connected pair of nodes $\{i, j\}\in E$, we construct two directed edges $i\to j$ and $j \to i$, with weights equal to the time $t_{i\to j}$ it would take for $i$ to transmit the disease to $j$, and the time $t_{j\to i}$ it would take for $j$ to transmit the disease to $i$, respectively. Both times are drawn from an exponential distribution with rate $\beta$, that is, $p(t_{i\to j})=\beta e^{-\beta t_{i\to j}}$. We also draw a random time $r_i$ it takes for each node $i \in V$ to recover from an exponential distribution with rate $\alpha$, that is, $p(r_i)=\alpha e^{- \alpha r_i}$. Suppose node $i$ becomes infected at time $t_i$. In the SIR dynamics, the infection is transmitted from node $i$ to node $j$ at time $t_j = t_i+t_{i\to j}$ if $i$ remains infected (i.e., $t_{i\to j}< r_i$) and $j$ remains susceptible by then. Eventually, $i$ recovers at time $t_i + r_i$.
In the SI dynamics, an infectious node remains infected forever, that is, $\alpha\to 0$. Thus, infection is transmitted from $i$ to $j$ at time $t_j = t_i+t_{i\to j}$ if $j$ remains susceptible.
Then, the epidemic percolation network (EPN) consists of only the directed edges $i\to j$ that transmit the infection (i.e., $t_{i \to j} < r_i$) with transmission time $t_{i\to j}$. Under this model, the transmission probability $T$, the probability of transmission between two nodes, can be written as~\cite{newman_spread_2002}
\begin{equation}
    T = \frac{\beta}{\alpha+\beta},
\end{equation}
where $T=1$ represents the SI dynamics.

\subsection{Reproduction number per generation}

In a partially susceptible population, the state of the epidemic is typically represented by an effective reproduction number. In this work, we quantify the effective reproduction number as a function of generation, denoted by $R_g$. Consider that infection begins at node $u_{0}$, randomly chosen from the set of nodes in the EPN. 
A susceptible node $i$ in the EPN becomes infected through a path $\gamma$, connecting $u_{0}$ to $i$. The transmission time of such a path in the EPN is the sum of the transmission times of its constituent edges, i.e.,
\begin{equation}
\sum_{ (i\to j) \in\gamma} t_{i\to j},
\end{equation}
where $t_{i\to j}$ denotes the transmission time of edge $i\to j$. We define the generation $g(i)$ of an infectious node $i$ as the number of edges of the path with the shortest transmission time from $u_0$ to $i$. In this framework, the initially infected node belongs to generation zero, $g(u_0) = 0$, and each node in generation $g+1$ is infected by a node in generation $g$. We define the reproduction number per generation, $R_g$, as the expected number of secondary infections caused by a node infected in generation $g$. Therefore, the reproduction number per generation is given by
\begin{equation}\label{eq: r_g}
    R_g=\frac{n_{g+1}}{n_g},
\end{equation}
where $n_g$ is the number of nodes in generation $g$. The value of $R_g$, as a function of $g$, determines the depth of the spread of infection. A larger value of $g_{\text{max}} = \max\{g \mid R_g > 0\}$ implies a deeper spread, meaning that on average the outbreak reaches susceptible nodes through longer transmission paths. 

For $g=0$, the reproduction number per generation is equal to the basic reproduction number $R_0$. The early phase of spread is commonly approximated by a branching process with a reproductive rate $R_0$.

\subsection{Network structure}

A contact network $G=(V, E)$ provides a representation of interactions as a set of edges $E$, connecting a set of nodes $V$, representing individuals. When information about interactions between individuals is limited, a common assumption is that they interact randomly, potentially with different levels of activity. In this scenario, the contact network can be modelled using the configuration model, with its degree distribution reflecting the heterogeneity in contact numbers.

However, real-world interactions are not necessarily random, but are often shaped by spatial and social constraints. For example, frequent physical contact typically occurs between individuals who live together or near each other, such as family members or roommates, or between individuals who share jobs or hobbies, such as coworkers and friends. In other words, the specific characteristics of the individuals (nodes) can influence the formation of connections (edges) in the network. In such cases, geometric networks provide a suitable framework to model contact networks. In geometric networks, nodes are assigned coordinates in a metric space, and edges are more likely to be formed between nodes closer in the metric space~\cite{boguna_network_2021, boguna_class_2003}. When modelling real-world networks with geometric network models, the coordinates assigned to nodes represent the characteristics that affect their connections.

Let us refer to the tendency to form connections based on proximity within an underlying metric space as spatiality. The level of spatiality differs between various network models. Configuration model networks, including Erd\H{o}s-R\'{e}nyi graphs, lack spatial structure. In contrast, random geometric graphs exhibit maximal spatiality.
In a random geometric graph, the nodes are uniformly distributed within a unit square in a $d$-dimensional Euclidean space. Here, we assume that our space is two-dimensional and that the nodes are distributed according to a Poisson point process with intensity $\rho$. An edge is formed between two nodes if their Euclidean distance is less than $\sqrt{\langle k\rangle / (\pi\rho)}$, where $\langle k\rangle$ is the average degree of the network. 
Both random geometric graphs and Erd\H{o}s-R\'{e}nyi graphs exhibit the Poisson degree distribution. Hence, any difference in dynamics in these two network models arises from the presence of spatial structure in random geometric graphs and the lack of spatiality in Erd\H{o}s-R\'{e}nyi graphs.

Furthermore, we use the heterogeneous spatial network (HSN) model, introduced by Boguñá et al.~\cite{boguna_small_2020}, to explore the behaviour of the reproduction number across the spatiality and degree heterogeneity spectrum. In the HSN model, each node is assigned an expected degree and coordinates in the metric space. Given these node characteristics, the model generates an undirected network where (i) the degree of each node is a Poisson random variable with a mean equal to the expected degree of the node, and (ii) the nodes are more likely to be linked with each other if they are close to each other in the metric space. The spatiality of the network, that is, the propensity of the nodes to connect to other nodes in their proximity, is governed by an independent parameter called the temperature, denoted by $\tau$. High values of $\tau$ correspond to the configuration model, where nodes form connections randomly regardless of their positions. In contrast, as $\tau\to 0$, nearby nodes are more likely to be connected, resulting in a strongly spatial network. We use this model to generate networks with negative binomial degree distributions with varying levels of spatiality controlled by $\tau$. The degree heterogeneity is parameterised by the dispersion parameter $r$ of the binomial distribution; a smaller $r$ implies a more heterogeneous degree distribution. 

\section{Results}\label{sec: results}

\subsection{The basic reproduction number}\label{sec: results_r_0}

In this section, we discuss the basic reproduction number, $R_0$, in Erd\H{o}s-R\'{e}nyi graphs and random geometric graphs. Recall that, in our setup, infection starts spreading from a single initially infected individual, $u_0$. Hence, the basic reproduction number is equal to the expected number of nodes in the first generation because every node in the first generation is necessarily infected by $u_0$. Moreover, each node in the first generation is a neighbor of $u_0$ in the EPN.

The basic reproduction number in an arbitrary network depends on the transmission probability, the expected node degree, and the presence of short loops.
Let us first discuss the case where the contact network is a tree, which contains no loops. Under this assumption, the EPN is also a tree. Since there is exactly one path between two nodes in a tree, all neighbors of $u_0$ in the EPN belong to the first generation and can only get infected through the edge from $u_0$. Therefore, the basic reproduction number in a tree is given by
\begin{equation}\label{eq:rnot_tree}
    R_0=T\langle k\rangle,
\end{equation}
where $\langle k \rangle$ is the average degree and $T$ is the transmission probability.

\begin{figure}
    \centering
    \includegraphics[width=\linewidth]{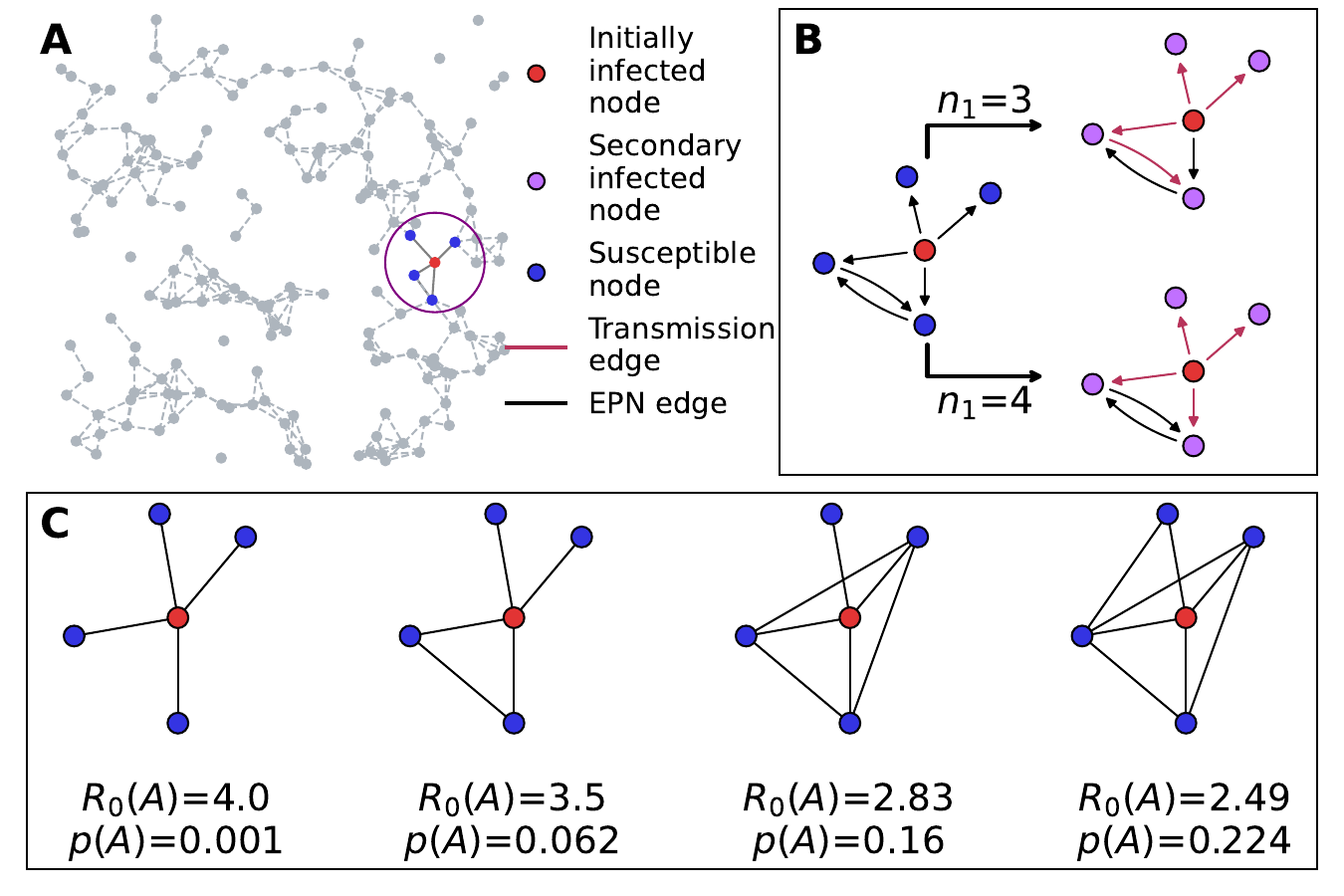}
    \caption{\textbf{The basic reproduction number in a random geometric graph.} 
    (A) Visualization of a random geometric graph. The nodes relevant for the computation of the basic reproduction number $R_0$, defined as the expected number of neighbors (in blue) infected directly by the initially infected node (in red), are shown in the circle.
    (B) In the presence of an EPN edge between neighbors, the same EPN configuration can lead to two different numbers of nodes in the first generation, $n_1$.
    (C) An initially infected node (in red) with the same degree can have different neighborhood isomorphic classes $A$, each occurring with probability $p(A)$ and resulting in the expected number of secondary cases, $R_0(A)$.
    }
    \label{fig:visualization_theory}
\end{figure}

In contrast, the presence of triangles results in a smaller basic reproduction number. 
If an edge exists between two neighbors of the initially infected node $u_0$, there are two distinct paths through which the infection can spread from $u_0$ to each of the two neighbors. The first path directly traverses the edge between $u_0$ and the neighbor. The second is a path of length two that goes from $u_0$ to the other neighbor and then traverses the edge between the neighbors. These two paths compete in transmission, and the infection will travel along the path with the shorter transmission time (Fig.~\ref{fig:visualization_theory}B). The susceptible neighbor will get infected through the edge with the least transmission time.

\begin{figure}
    \centering
    \includegraphics[width=\linewidth]{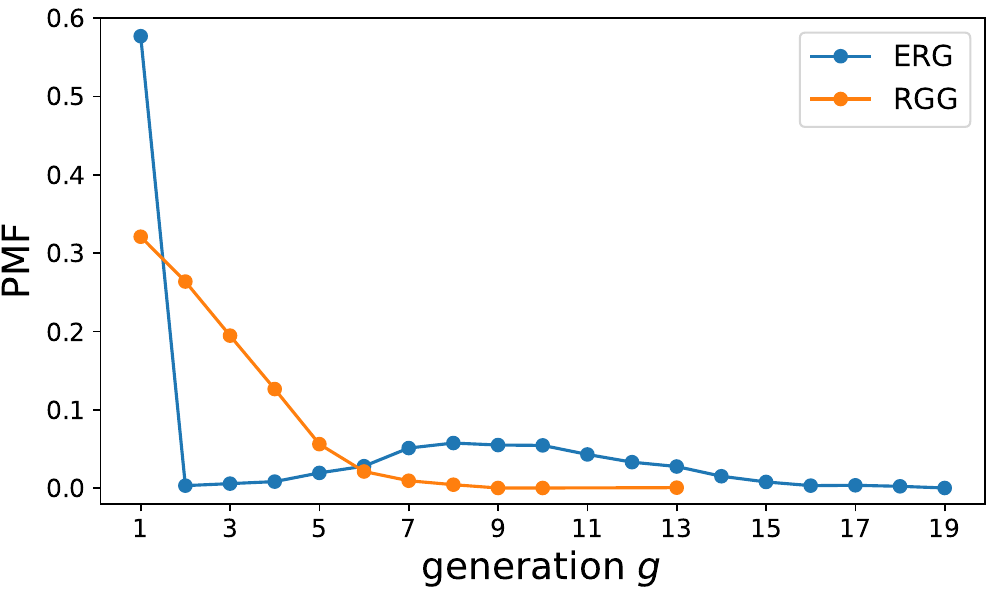}
    \caption{\textbf{The distribution of the generation $g$ of the neighbors of the initially infected node in Erd\H{o}s-R\'{e}nyi graphs (ERGs) and random geometric graphs (RGGs).} The network size is $N=10^4$ and the average degree is $\langle k\rangle =12$. The transmission rate is $\beta=1$, and the recovery rate is $\alpha=0.001$. The result is an average over $200$ runs.}
    \label{fig: r_not_rgg}
\end{figure}

\begin{figure}
    \centering
    \includegraphics[width=\linewidth]{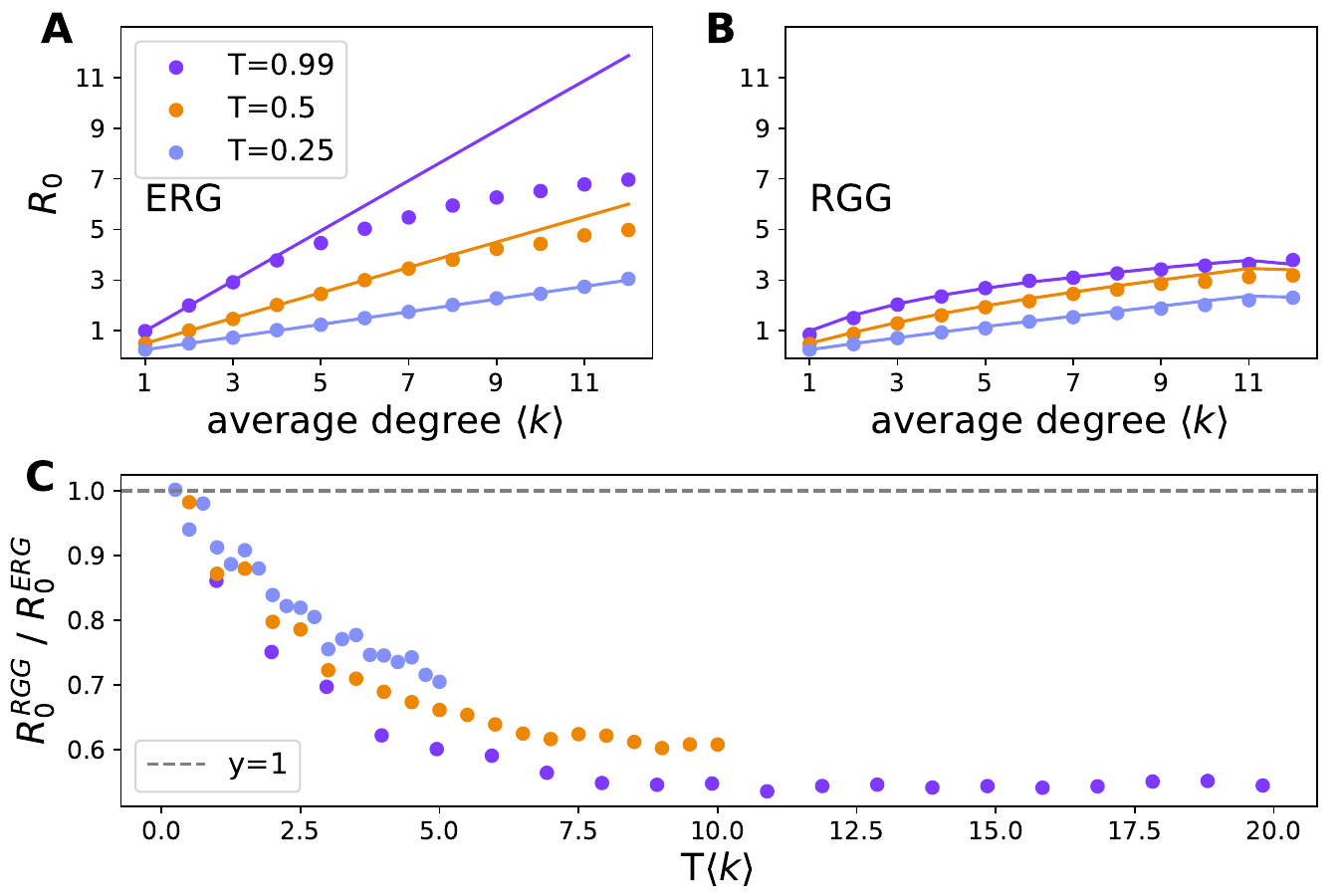}
    \caption{\textbf{ The basic reproduction number $R_0$ as a function of average degree $\langle k \rangle$ and transmission probability $T$.} A) Shows $R_0$ as a function of $\langle k \rangle$ for Erd\H{o}s-R\'{e}nyi graphs (ERGs) and B) for random geometric graphs (RGGs). C) Shows the ratio between $R_0$ in RGG to $R_0$ in ERG as a function of $T\langle  k \rangle$. The transmission probability is $T\in\{0.25, 0.5, 0.99\}$. Each simulation and analytical result (points and lines, respectively) is an average over $500$ runs on a network of size $N=10000$. For ERGs, the analytical lines follow the tree-like assumption of Eq.~\eqref{eq:rnot_tree}. For RGGs, the analytical lines are calculated from Eq.~\eqref{eq: r_0_rgg}.}
    \label{fig: r_not_erg}
\end{figure}

Figure~\ref{fig: r_not_rgg} shows the probability distribution for the generation of the neighbors of the initially infected node in the EPN. As this figure shows, in an Erd\H{o}s-R\'{e}nyi graph, the only way a susceptible neighbor is not directly infected by the initially infected node is when the transmission travels through a long cyclic path and reaches the neighbor before direct transmission occurs. Figure~\ref{fig: r_not_erg}A compares the value of $R_0$ for Erd\H{o}s-R\'{e}nyi graphs with different expected degree values with their equivalent in a tree, Eq.~\eqref{eq:rnot_tree}. For a small expected degree, Eq.~\eqref{eq:rnot_tree} provides an accurate approximation 

In random geometric graphs, however, triangles in the EPN are not negligible. As Fig.~\ref{fig: r_not_rgg} shows, a significant fraction of neighbors belong to the second and third generations, meaning that they have been infected through paths of length two and three. Nodes in random geometric graphs are connected based on proximity, which increases the frequency and influence of short loops compared to long loops as pathways for infecting the neighbors of the initially infectious individual. In particular, loops of length three---the shortest possible---play a crucial role in determining the basic reproduction number.

In the following, we present a more precise approximation of $R_0$ in random geometric graphs by explicitly accounting for loops of length three. This approximation, which we elaborate in the following, is in good agreement with simulation results, as shown in Fig.~\ref{fig: r_not_erg}B. 

The initially infected node $u_0$ can transmit the infection to its neighbors until it recovers. The basic reproduction number, $R_0$, equals the expected number of neighbors infected directly from a randomly chosen $u_0$. Let us focus on the subgraph of the contact network induced by $u_0$ and the set of its neighbors, denoted by $\mathcal{N}(u_0)$ (Fig.~\ref{fig:visualization_theory}A). For simplicity, we relabel the initially infected node as $u_0=0$ and other nodes in this subgraph as $i\in \mathcal{N}(0)$ for $i=1,\dots, k$, where $k=|\mathcal{N}(0)|$ is the degree of node $0$. We represent the structure of the subgraph induced by the nodes in set $\{0\}\cup \mathcal{N}(0)$ by an adjacency matrix $A$, where the first row/column corresponds to the initially infected node. Then, $A_{0i}=1$ for all $i\in \mathcal{N}(0)$ and $A_{ij}=1$ if $i$ and $j$ are connected. 

In this setup, the basic reproduction number can be computed as 
\begin{equation}
    R_0 = \sum_{A} R_0(A) p(A),
\end{equation}
where $p(A)$ denotes the probability of configuration $A$ and $R_0(A)$ is the basic reproduction number for a specific configuration $A$ around node $0$.

We calculate $p(A)$ by generating random geometric graphs with $N$ nodes and expected degree $\langle k \rangle$, and repeatedly sampling the neighborhood of a randomly chosen node. By counting the times each $A$ appears, we can find the ensemble of all adjacency matrices $A$ in a network with a certain average degree $\langle k \rangle$ and calculate $p(A)$ (Fig.~\ref{fig:visualization_theory}C).

To calculate $R_0(A)$, we make the simplifying assumption that nodes in $\mathcal{N}(0)$ become infected either through paths of length one (single edges) or two from node $0$. Hence, node $i \in \mathcal{N}(0)$ will remain susceptible, or get infected either by node $0$ through the path $0\to i$ or by an already infected neighbor $j\in (\mathcal{N}(0)\cap \mathcal{N}(i))$ through the path $0\to j\to i$. The number of paths of length two from node $0$ to its neighbor $i$ is given by
\begin{equation}
    A^2_{0i} = \sum_{j=0}^{k}A_{0j} A_{ji},
\end{equation}
and uniquely determined by $A$.

Node $i$ can be infected through its common neighbor with node $0$ that was infected earlier. We represent the set of such neighbors by $V_i$

\begin{equation}
    V_i = \{j\mid j\in(\mathcal{N}(0)\cap \mathcal{N}(i)),\; t_{0\to j}<t_{0\to i}\}.
\end{equation}

In the following, we proceed in four steps. First, we compute the probability distribution of the cardinality $|V_i|$, which represents the number of competitors of the initially infected node $0$ attempting to infect node $i$. Second, we determine the probability that the transmission time from node $0$ to node $i$ is shorter than the recovery time of node $0$. Third, we evaluate the probability that node $0$ infects node $i$ before any other node in $V_i$. Finally, by combining the results of the previous three steps for all $i\in \mathcal{N}(0)$, we derive $R_0(A)$. 

By letting $\sigma$ denote the permutation on $\mathcal{N}(0)$, where $\sigma(j)<\sigma(i)$ indicates $t_{0\to j}<t_{0\to i}$, we can write the probability of the cardinality of set $V_i$ as
\begin{equation}
\begin{aligned}
&p(|V_i|=m\mid A^2_{0i}, \sigma(i)=\pi)\\
&=\frac{1}{\binom{k-1}{A^2_{0i}}}\binom{\pi-1}{m}\binom{k-\pi}{A^2_{0i}-m}.
\end{aligned}
\end{equation}

Next, the probability that the initial node $0$ can transmit the disease to $i$ before recovering, conditioned on $\sigma(i)$, is 
\begin{equation}
\begin{aligned}
    &p(t_{0\to i}< r_0\mid\sigma(i)=\pi)
    =\frac{\sum_{\sigma:\sigma(i)=\pi}p(t_{0\to i}<r_0, \sigma)}{p(\pi)}\\
    ={}&k!\int_{0}^{\infty}dx_1\beta e^{-\beta x_1}\dots\int_{x_{\sigma(i)-1}}^{\infty}dx_{\pi} \beta e^{-\beta x_{\pi}} \int_{x_{\pi}}^{\infty}dr \alpha e^{-\alpha r}  \\
    &\times \int_{x_{\pi}}^{\infty}dx_{\pi_i+1}\beta e^{-\beta x_{\pi+1}}\dots\int_{x_{k-1}}^{\infty}dx_k\beta e^{-\beta x_{k}}  \\
    ={}&\frac{k!}{(k-\pi)!} \prod_{j=1}^{\pi}\frac{\beta}{\alpha+(k-\pi+j)\beta}.
\end{aligned}
\end{equation}

Finally, to complete the chain of conditional probabilities, we assume that the transmission from node $0$ to node $i$ takes place before node $0$ recovers, i.e., $t_{0\to i}<r_0$. This implies $t_{0\to j}<r_0$ for all $j\in V_i$.
To infect node $i$, node $0$ competes with all neighbors of $i$ in $V_i$ that can transmit the infection to $i$ before they recover. Then, the probability that transmission takes place through the path $0\to i$, conditioned on $t_{0\to i}<r_0$ and $|V_i|$, is
\begin{equation}
\begin{aligned}
    &p(\mathbf{1}_{0\to i}=1 \mid t_{0\to i}<r_0, |V_i|=m) \\&=\sum_{x=0}^{m}\binom{m}{x}\frac{T^{m}(1-T)^{m-x}}{x+1},
\end{aligned}
\end{equation}
where $T = \beta / (\alpha+\beta)$ is the transmission probability, and the indicator function $\mathbf{1}_{0\to i}$ is defined as
\begin{equation}
    \mathbf{1}_{0\to i}=\begin{cases}
        1&\text{if transmission happens through edge $(0, i)$}\\
        0& \text{otherwise}.
    \end{cases}
\end{equation}

The probability $R(\mathbf{1}_{0\to i}=1\mid\sigma(i)=\pi, A^2_{0i})$ that a node $i\in \mathcal{N}(u_0)$ gets infected through the path $0\to i$, conditioned on $\sigma$ and $A$, is
\begin{equation}
\begin{aligned}
    &R(\mathbf{1}_{0\to i}=1\mid\sigma(i)=\pi, A^2_{0i})\\
    ={}&p(t_{0\to i}<r_0\mid \sigma(i)=\pi)\\ &\times\sum_{m=0}^{A^2_{0i}} p(\mathbf{1}_{0\to i}=1\mid t_{0\to i}<r_0,|V_i|=m)\\ &\times p(|V_i|=m\mid A^2_{0i},\sigma(i)=\pi).
\end{aligned}
\end{equation}
Summing over all values of $\sigma(i)$ and all nodes $i$, $R_0(A)$ is 
\begin{equation}\label{eq: r_0_rgg}
\begin{aligned}
    R_0(A)={}&\sum_{i=1}^{k}\sum_{\pi=1}^k \frac{1}{k}R(\mathbf{1}_{0\to i}=1\mid\sigma(i)=\pi, A^2_{0i}).
    \end{aligned}
\end{equation}

The SI dynamic is a special case where $T=1$. Then, $R_0(A)$ can be written as
\begin{equation}
\begin{aligned}
    R_0(A)={}&\sum_{i=1}^k\frac{1}{1+A^2_{0i}}\sum_{l=0}^{A^2_{0i}} \frac{1}{1+l}.
\end{aligned}
\end{equation}

Figure~\ref{fig: r_not_erg}C compares the ratio of the basic reproduction number $R_0$ in random geometric graphs to $R_0$ in Erd\H{o}s-R\'{e}nyi graphs as a function of $T\langle k\rangle$. As the transmission probability and the average degree increase, the difference in $R_0$ between the random geometric graph and the Erd\H{o}s-R\'{e}nyi graph becomes more pronounced, with $R_0$ in the random geometric graph approaching approximately half of the value of $R_0$ in the Erd\H{o}s-R\'{e}nyi graph. It is worth highlighting that this is due to the spatiality alone, as the average degree, the degree distribution, the transmission probability, and the network size are the same for both graphs. 

While the value of $R_0$ in both Erd\H{o}s-R\'{e}nyi graphs and random geometric graphs deviates from that of a tree with the same average degree and degree heterogeneity, this deviation is more pronounced in random geometric graphs (Fig.~\ref{fig: r_not_erg}). The main difference in $R_0$ between Erd\H{o}s-R\'{e}nyi graphs and random geometric graphs arises from the contribution of loops of various lengths in transmitting the infection (Fig.~\ref{fig: r_not_rgg}). Although short loops are negligible in Erd\H{o}s-R\'{e}nyi graphs, these graphs contain a significant number of long loops. However, since transmission along a long path is, on average, slower than transmission across a single edge, the probability that it competes with direct transmission from the initially infected node to its neighbor is small, resulting in a smaller deviation from tree-like behavior.

\subsection{The reproduction number per generation}\label{sec: results_r_g}

In this section, we will derive the reproduction number per generation $R_g$ for Erd\H{o}s-R\'{e}nyi graphs and random geometric graphs. We first focus on SI dynamics, modelled by first passage percolation. For Erd\H{o}s-R\'{e}nyi graph, we will base our argument on the branching process on finite graphs, and for random geometric graphs, we show how the spatial constraint on network formation will affect the reproduction chance. Then, we study the SIR dynamics by mapping it to a special case of first passage percolation.

\begin{figure}
    \centering
    \includegraphics[width=\linewidth]{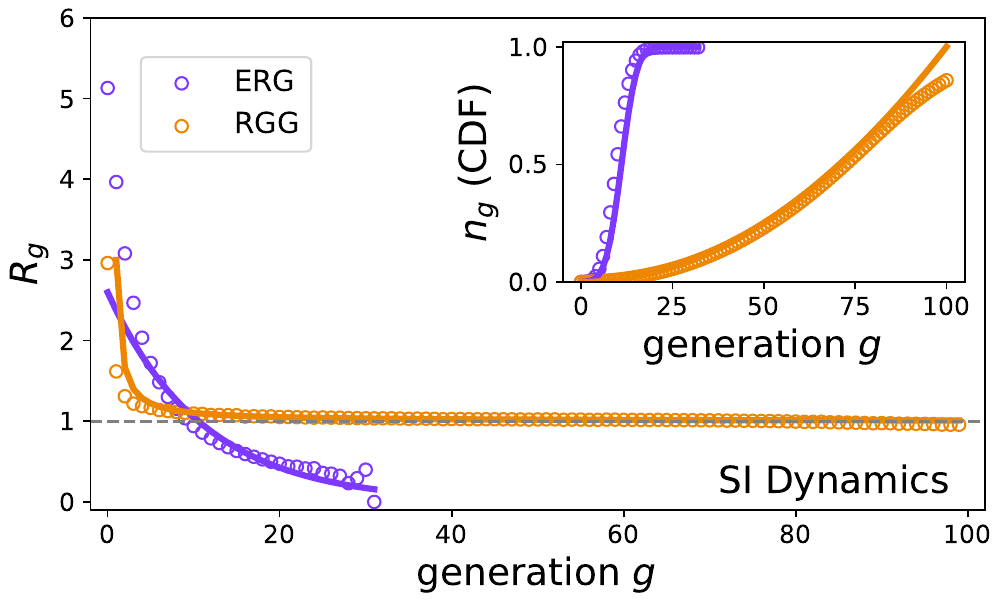}
    \caption{\textbf{The reproduction number per generation $R_g$.} Results are shown for the SI dynamics on Erd\H{o}s-R\'{e}nyi graphs (ERGs) and random geometric graphs (RGGs). The inner plot illustrates the normalized number of nodes per generation $n_g$. Points indicate simulations and lines show analytical results. All networks have size $N=10^4$ and average degree $\langle k\rangle=6$. Results are averaged over $500$ runs.}
    \label{fig: r_g}
\end{figure}

As defined in Eq.~\eqref{eq: r_g}, $R_g$ is determined by the number of nodes in each generation, $n_g$. To derive $n_g$ for Erd\H{o}s-R\'{e}nyi graphs, we present a theory introduced by Bhamidi et al.~\cite{bhamidi_first_2010}.
Consider first-passage percolation, where the transmission times of edges $t_{i\to j}$ are independent and identically distributed random variables with an exponential distribution $p(t_{i\to j})=e^{-t_{i\to j}}$. Theorem \ref{th: conf_r_g} characterizes the probability distribution for the generation of an infected individual in network $G$.
\begin{theorem} \label{th: conf_r_g}
    Let $G$ be an Erd\H{o}s-R\'{e}nyi graph with $n$ nodes and average degree $\langle k\rangle>1$. Then $p(g)$, the probability that an infected individual belongs to generation $g$, is as below
    \begin{equation}\label{eq: n_g_erg}
        p(g) = N\left(a\log(n),a\log(n)\right),
    \end{equation}
    where $N$ is the standard normal distribution, and 
    \begin{equation}
        a=\frac{\langle k\rangle}{\langle k\rangle-1}.
    \end{equation}
\end{theorem}
Since the reproduction number per generation,$R_g$, is defined as the ratio between the numbers of individuals in consecutive generations as in Eq.~\eqref{eq: r_g}, we have 

\begin{equation}\label{eq: r_g_erg}
    R_g = \frac{p(g+1)}{p(g)} = \exp\left[1-\frac{1+2g}{2a \log n}\right].
\end{equation}
Hence, in Erd\H{o}s-R\'{e}nyi graphs, $R_g$ decays exponentially until the percolation process halts, as illustrated in Fig.~\ref{fig: r_g}. This rapid decline results in a small $g_{max}$.

We now proceed to calculating $R_g$ for SI dynamics in random geometric graphs, where our strategy is again to derive $n_g$. For this, we first calculate the average time of infection per generation. Node $i$ which belongs to generation $g$ has at least one neighbor from generation $g-1$. Consider the set of random variables $X_i^g=\{t_{j\to i}\mid j\in \mathcal{N}(i), g(j)=g-1, t_j<t_i\}$ where $t_{i\to j}$ is the transmission time along the edge $i\to j$. Define $Y^g_i = \min X_i^g$ as the time difference between the infection of node $i$ and the node that infects it. By definition, the cumulative distribution function of $Y_i^g$ is given by
\begin{equation}
    F_{Y_i^g}(y)=P(Y_i^g\leq y)=1-\prod_{x\in X_i^g} P(x>y)=1-e^{-\beta\lvert X_i^g \rvert y}
\end{equation}
where $\lvert X_i^g \rvert$ is the cardinality of the set $X_i^g$. The last equality holds since $t_{i\to j}$ follows an exponential distribution, $P(x>y)=e^{-\beta y}$.
The expected value of $Y_i^g$ is
\begin{equation}
    \mathbb{E}[Y_i^g] = \frac{1}{\beta \lvert X_i^g \rvert}.
\end{equation}
The average of $\mathbb{E}[Y_i^g]$ over all nodes in generation $g$, denoted by $\mathbb{E}[Y^g]$, is
\begin{equation}
    \mathbb{E}[Y^g] = \frac{1}{\beta}\mathbb{E}\left[\frac{1}{\lvert X_i^g \rvert}\right]=\frac{1}{\beta m_g},
\end{equation}
where we introduce
\begin{equation}
    m_g = \mathbb{E}\left[\frac{1}{\lvert X_i^g \rvert}\right]^{-1}.
\end{equation}

\begin{figure}[!h]
    \centering
    \includegraphics[width=\linewidth]{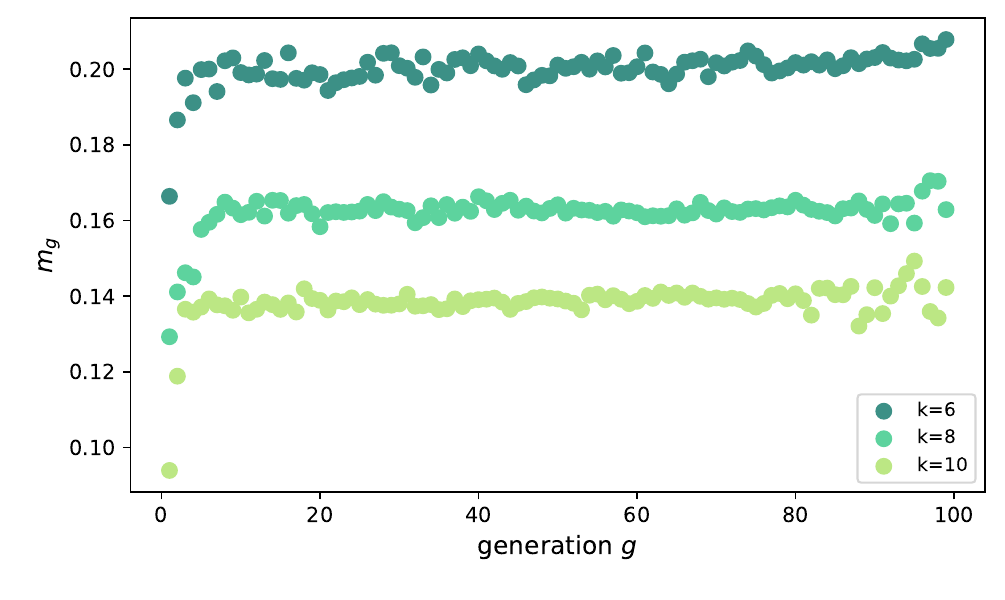}
    \caption{\textbf{Average fraction of neighbors infected in the generation before the node.} Results are shown for $T=0.99$. The networks are random geometric graphs. All networks have
size $N = 10^4$. Results are averaged over $200$ simulation runs.}
    \label{fig:m_g_plot}
\end{figure}

We assume $m_g$ is a function of transmission rate $\beta$ and average degree $\langle k\rangle$ and does not depend on generation, i.e., $m_g  = \bar{m}(\langle k\rangle, \beta)$. This is empirically validated as shown in SI Fig.~\ref{fig:m_g_plot}. Hence,
\begin{equation}\label{eq: E_Y}
    \mathbb{E}[Y^g]=\frac{1}{\beta \bar{m}(\langle k\rangle, \beta)}.
\end{equation}
Let $t_g$ denote the expected time of infection for a node in generation $g$. Then,
\begin{equation}\label{eq: t_g_rec}
    t_g = t_{g-1} + \mathbb{E}[Y^g].
\end{equation}
Combining Eqs.~\eqref{eq: E_Y} and \eqref{eq: t_g_rec}, we can conclude
\begin{equation}\label{eq: t_g}
    t_g =  \sum_{x=1}^g \mathbb{E}[Y^g]=\frac{g}{ \langle k\rangle\bar{m}(\langle k\rangle, \beta)}.
\end{equation}
Equation~\eqref{eq: t_g} indicates that the expected time of infection per generation $t_g$ is linearly proportional to the generation $g$.

The next step is to find the number of nodes infected as a function of time.
We locate the initially infected node $u_{0}$ at the origin $o$, and $\upsilon_d$ denotes the volume of the unit ball in $\mathbb{R}^d$. Define $H_t$ as the subset of nodes reached by the SI dynamics starting from the origin by the time $t$. Let $B(x, s)$ be the Euclidean ball centred at $x$ with radius $s$. Theorem 1.1 in Coletti et al.~\cite{coletti_limiting_2023} indicates that there exists a constant $\phi > 0$ such that for any $\varepsilon \in (0,1)$, almost surely for large $t$:
\begin{equation}\label{eq: h_t}
    (1 - \varepsilon) B(o, \phi) \subseteq \frac{1}{t} H_t \subseteq (1 + \varepsilon) B(o, \phi).
\end{equation}
Combining Eqs.~\eqref{eq: t_g} and \eqref{eq: h_t}, we can conclude that
 \begin{equation}\label{eq:n_g_rgg}
     \sum_{x=0}^{g} n_x  = C g^2,
 \end{equation}
where $C$ is a constant related to the node density. Now, we can calculate the reproduction number as
 \begin{equation}\label{eq: r_g_rgg}
     R_g =\frac{2g+1}{2g-1}.
 \end{equation}
Hence, as $g$ increases, $R_g$ converges to unity, in agreement with the simulation results shown in Fig.~\ref{fig: r_g}.

In SIR dynamics, each edge $(i, j)$ can transmit the disease from $i$ to $j$ if $t_{ij}<r_i$. The probability of transmission time $t_{ij}$, conditioned on $t_{ij}<r_i$ is
\begin{equation}
    P(t_{ij}|t_{ij}<r_i)=\frac{P(t_{ij},t_{ij}<r_i)}{P(t_{ij}<r_i)},
\end{equation}
where
\begin{equation}
    P(t_{ij}<r_i)=\frac{\beta}{\alpha+\beta},
\end{equation}
and
\begin{equation}
    P(t_{ij},t_{ij}<r_i)=\beta e^{-\beta t_{ij}}\int_{t_{ij}}^\infty \alpha e^{-\alpha x} dx=\beta e^{-(\alpha + \beta)t_{ij}}.
\end{equation}
Hence,
\begin{equation}
    P(t_{ij}|t_{ij}<r_i)= (\alpha +\beta) e^{-(\alpha + \beta)t_{ij}}.
\end{equation}
The transmission time of edges in the EPN still follows the exponential distribution, but with rate $\alpha+\beta$. As a result, the reproduction number per generation under the SIR dynamics follows Eqs.~\eqref{eq: r_g_erg} and~\eqref{eq: r_g_rgg} for Erd\H{o}s-R\'{e}nyi graphs and random geometric graphs, respectively.

Figure~\ref{fig: r_g} illustrates the reproduction number $R_g$ as a function of generation $g$ for Erd\H{o}s-R\'{e}nyi graphs and random geometric graphs. The inset plot shows that the $n_g$ results from the simulations of both Erd\H{o}s-R\'{e}nyi graphs and random geometric graphs are in agreement with the analytical results presented in Eqs.~\eqref{eq: n_g_erg} and~\eqref{eq:n_g_rgg}, respectively. In Erd\H{o}s-R\'{e}nyi graphs, $R_g$ decays exponentially until the outbreak ends due to limited network size. In a random geometric graph, the decay in $R_g$ is faster than Erd\H{o}s-R\'{e}nyi graphs in the beginning until $R_g$ reaches unity. As $n_g$ grows cumulatively with $g^2$, $R_g$ remains equal to unity until percolation halts and no more Infected nodes are produced.

\begin{figure}[!h]
    \centering
    \includegraphics[width=\linewidth]{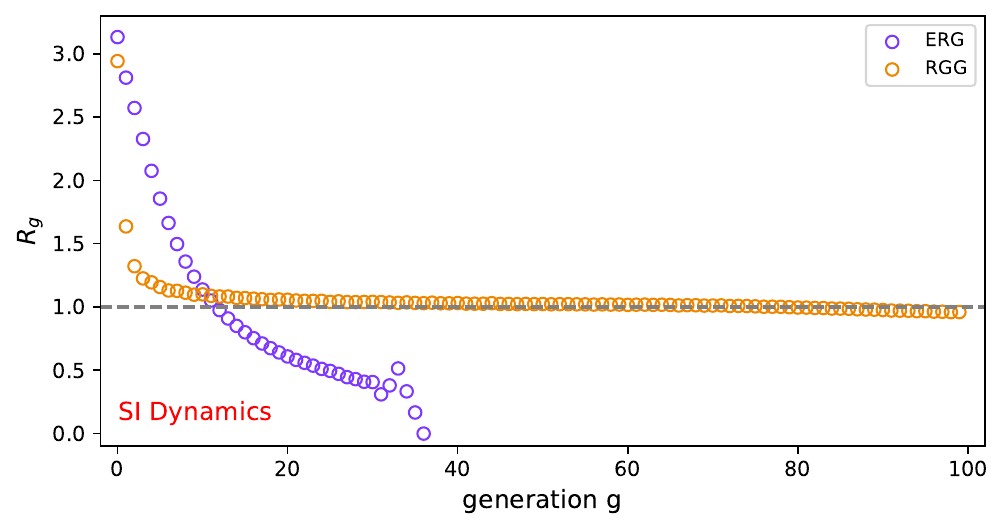}
    \caption{\textbf{The reproduction number per generation $R_g$ for networks with the same basic reproduction number $R_0$.} Results are shown for the SI dynamics on Erd\H{o}s-R\'{e}nyi graphs (ERGs) and random geometric graphs (RGGs). All networks have size $N=10^4$ and the basic reproduction number $R_0\approx 3$. The average degree of the ERG is $\langle k\rangle=3$ and the average degree of RGG is $\langle k\rangle=6$. Results are averaged over $500$ runs.}
    \label{fig: r_g_same_r0}
\end{figure}

It is worth highlighting that the difference in Erd\H{o}s-R\'{e}nyi graphs and random geometric graphs in terms of $R_g$ is due to the spatial network structure and is not caused by different $R_0$ values (SI Fig.~\ref{fig: r_g_same_r0}).

\subsection{The reproduction number $R_{g}$ is determined by the competition between neighbors}

In this section, we explain how the reproduction number per generation can be concluded from the competition between infected nodes in each generation, regardless of the network structure.

Similarly to our argument for the basic reproduction number, let us first consider the case where the contact network is formed by a branching process that results in a tree. In such a tree, since there are no loops, when a node becomes infected, it has only one infected neighbor: the node that transmitted the disease to it. 
Therefore, $R_g$ remains equal to $R_0$ until it drops to zero. 

However, in a network with loops, when a susceptible node becomes infected, there may be more than one neighbor that was infected earlier. In that case, these infected neighbors will compete to transmit the disease to the susceptible node. If a susceptible node has more infected neighbors at the time of its infection, the competition to infect it is stronger, and the chance of transmission for each infected individual is lower. 
Because of this competition, as the number of infected nodes increases, the reproduction number decreases from $R_0$, in contrast to the tree case.

\begin{figure*}
    \centering
    \includegraphics[width=0.8\linewidth]{
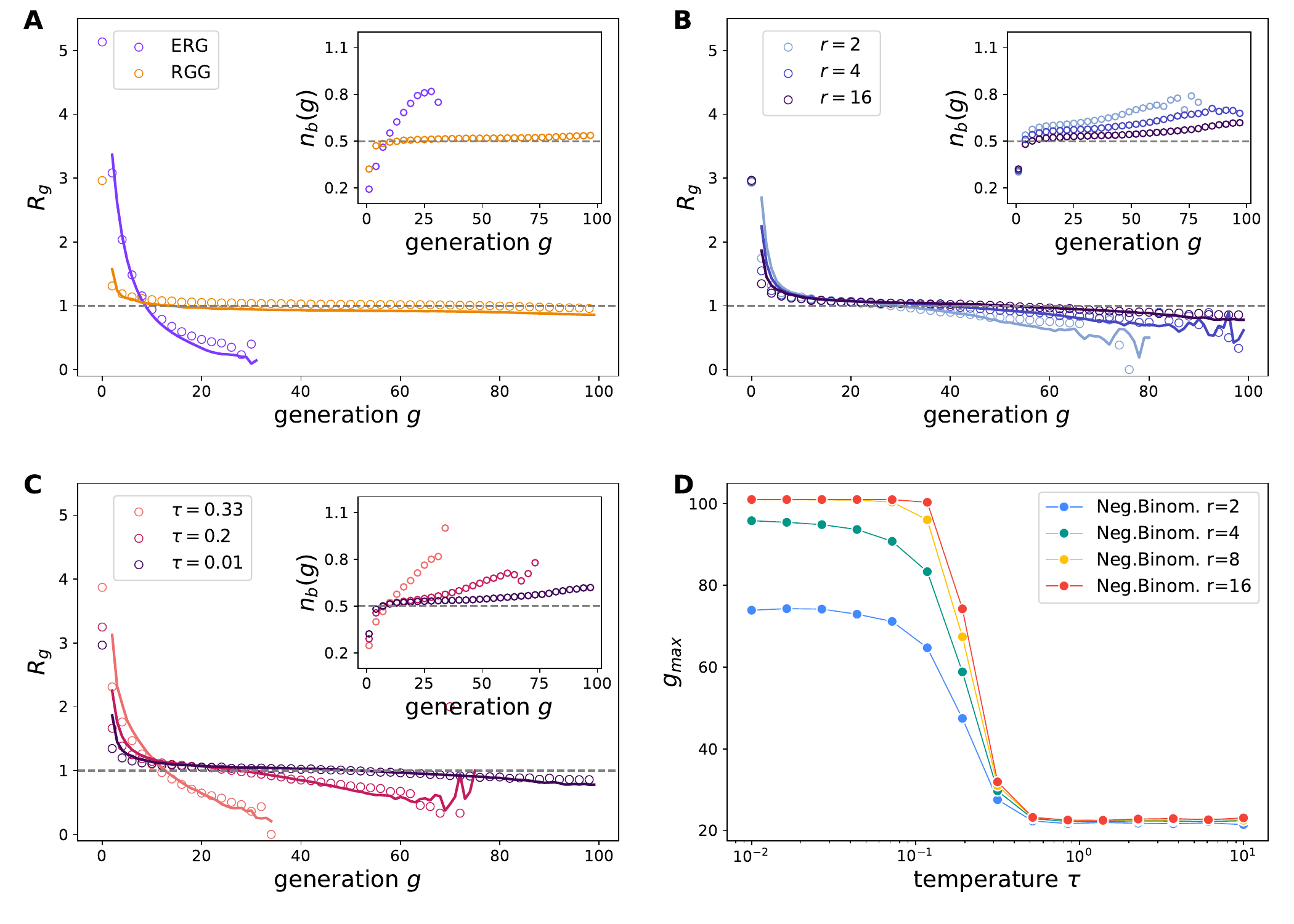}
    \caption{\textbf{
    The reproduction number per generation $R_g$ and competition between infected nodes.} (A-C) Points indicate $R_g$, and lines correspond to the right-hand side of Eq.~\eqref{eq: n_g_r_g} as a function of generation $g$. Insets show $n_\mathrm{b}(g)$. A) Results for Erd\H{o}s-R\'{e}nyi graphs (ERGs) and random geometric graphs (RGGs). B) Results for HSNs with negative binomial degree distributions with dispersion parameter $r\in[2, 4, 16]$ and spatiality $\tau=0.01$. C) Results for HSNs with negative binomial degree distributions with dispersion parameter $r=16$ and spatiality $\tau\in [0.33, 0.2, 0.01]$. D) Shows the total number of generations $g_{max}=\max\{g|R_g>0\}$. Results are shown for the temperatures $\tau\in(10^{-2}, 10)$ and the dispersion parameters $r\in[2, 4, 8, 16]$.  All Results correspond to SI dynamics. All HSNs have size $N=5000$, while ERGs and RGGs have size $N=10^4$. All networks have average degree $\langle k\rangle=6$. Results are averaged over $10000$ simulation runs.}
    \label{fig: n_g_het}
\end{figure*}

Figure~\ref{fig: n_g_het}A shows 
the average fraction of each node's neighbors infected earlier than the node, denoted by $n_\mathrm{b}(g)$, as a function of the generation $g$ of the node for Erd\H{o}s-R\'{e}nyi graphs and random geometric graphs. In an Erd\H{o}s-R\'{e}nyi graph, the fraction of infected neighbors of a susceptible node increases rapidly. This is because, as the population of infected nodes becomes larger, the chances of neighbors being infected increase. On the other hand, in random geometric graphs, this fraction increases more quickly with the generation, until it reaches the value of 0.5, i.e., half of the neighbors of a susceptible node are infected, after which it remains constant.

The fraction $n_\mathrm{b}(g)$ determines the value of $R_g$. The number of secondary infected individuals for a node in generation $g$ depends on the number of susceptible neighbors of the node at the time of its infection and the competition it faces when attempting to infect them. 
Since a node in generation $g+1$ gets infected through an edge leading to a node in generation $g$, we can conclude that a node in generation $g$ will infect each of its neighbors with probability $1/\left(\langle k_{g+1}\rangle n_\mathrm{b}(g+1)\right)$, where $\langle k_{g}\rangle $ is the average degree of a node in generation $g$. A node in generation $g$ has on average $\langle k_{g}\rangle (1-n_\mathrm{b}(g))$ susceptible neighbors when it gets infected and starts spreading the infection. Therefore, $R_g$, the expected number of secondary infections caused by a node infected in generation $g$, is
\begin{equation}\label{eq: n_g_r_g}
    R_g = \frac{\langle k_{g} \rangle (1-n_\mathrm{b}(g))}{\langle k_{g+1} \rangle n_\mathrm{b}(g+1)}.
\end{equation} 
Figure~\ref{fig: n_g_het}A shows the relationship between $R_g$ (scatter plot) and $\langle k_{g} \rangle (1-n_\mathrm{b}(g))/ \left(\langle k_{g+1} \rangle n_\mathrm{b}](g+1)\right)$ (line plot) as a function of generation $g$ for Erd\H{o}s-R\'{e}nyi graphs and random geometric graphs.
In Erd\H{o}s-R\'{e}nyi graphs, $n_\mathrm{b}(g)$ exponentially increases, as a result, $R_g$ decays exponentially. In a random geometric graph, however, $n_\mathrm{b}(g)$ first increases rapidly and then remains constant at $n_\mathrm{b}(g)\approx 0.5$ until the epidemic dies out. Hence
\begin{equation}
    R_g=\frac{1}{n_\mathrm{b}}-1=1 \quad \text{for large enough } g.
\end{equation}
Here, we use the fact that $\langle k_{g-1} \rangle\approx \langle k_{g} \rangle$ in random geometric graphs.

This relationship between $R_g$ and $n_b$ generalizes to a wider range of networks beyond Erd\H{o}s-R\'{e}nyi graphs and random geometric graphs. We validate this across the full spectrum of spatiality and degree heterogeneity using the heterogeneous spatial network (HSN) model. 
Figure~\ref{fig: n_g_het}B shows the relationship between the two sides of Eq.~\eqref{eq: n_g_r_g} as a function of generation $g$. Regardless of degree heterogeneity determined by $r$, in a network with spatial structure, $R_g$ goes to unity for large enough $g$, a behavior fully explained by the competition between infected neighbors at the time of infection. 
As Fig.~\ref{fig: n_g_het}C shows, the decrease of $R_g$ from unity towards zero becomes more gradual as the temperature $\tau$ decreases, i.e., spatiality increases.

Figure~\ref{fig: n_g_het}D shows the maximum generation $g_{max}$ for SI dynamics as a function of the dispersion parameter $r$ and the temperature $\tau$. Higher $g_{max}$ indicates a slower decay in the tail of $R_g$, as we observed for random geometric graphs. Moving from the random to spatial networks (high temperature to low temperature), the $g_{max}$ increases significantly around $\tau=0.3$, and the decay in $R_g$ becomes slower until $\lim_{g\to \infty}R_g=1$ as the network becomes spatial.

\section{Conclusion}\label{sec: conc}

In this paper, we have studied the effect of spatially embedded contact patterns on the basic and effective reproduction numbers. 
We represent the spatial structure of the population using network epidemic models. 
To facilitate our analysis, we introduced the reproduction number per generation, $R_g$, defined as the expected number of new infections caused by nodes in generation $g$. 

With a random contact structure, as captured by Erd\H{o}s-R\'{e}nyi graphs, the basic reproduction number can be approximated by the size of the first generation in a branching process with intensity proportional to the average degree of the network and the transmission probability. However, when the contacts are spatially constrained, as captured by random geometric graphs, we observed that the basic reproduction number $R_0$ is lower than predicted by the branching process with the same intensity (Figure~\ref{fig: r_not_erg}). This is because the spatial contact structure leads to the formation of triangles and other short loops, reducing the likelihood of direct disease transmission from the initially infected individual to its neighbors. To account for the impact of short loops in random geometric graphs, we included loops of length three (triangles) in our calculation of $R_0$. Comparing our analytical results with simulations demonstrates that incorporating these loops can substantially improve the accuracy of the $R_0$ estimate. 

As infection spreads in an Erd\H{o}s-R\'{e}nyi graph where there are no spatial constraints, the reproduction number per generation, $R_g$, decreases exponentially until it reaches zero, marking the end of the spreading. However, in the spatial case modelled by random geometric graphs, this decline is faster than exponential until $ R_g\approx 1$ (Figure~\ref{fig: r_g}). At that point, $R_g$ remains approximately constant until the epidemic dies out and no further generations are produced. This behavior is confirmed analytically by examining the number of nodes in each generation. In Erd\H{o}s-R\'{e}nyi graphs, the probability of an infected node belonging to each generation follows a normal distribution (Eq.~\ref{eq: n_g_erg}). In contrast, in random geometric graphs, the number of nodes reached by percolation up to a given generation grows quadratically with the number of generations (Eq.~\ref{eq: h_t}).

We have shown that $R_g$ is determined by 
the levels of competition between infected nodes to transmit the disease to susceptible nodes in generations $g$ and $g+1$, independently of the underlying network structure. This competition is caused by the presence of loops in the network, and can be quantified by the fraction of neighbors infected earlier, $n_\mathrm{b}$.

We have shown that increasing network spatiality, which increases the amount of loops, shifts the behavior of $R_g$ from an exponentially decreasing regime to one in which $R_g$ converges to $R_g\approx 1$ (Figure~\ref{fig: n_g_het}). We note that $n_\mathrm{b}(g)$ is related to the notion of force of infection in epidemiology. Thus, the asymptotic value of $R_g=1$ arises from a generation-invariant force of infection.

Our findings raise questions about the conventional use of reproduction numbers as a tool for identifying contagiousness and predicting the epidemic dynamics. In a homogeneously mixed population, the contagiousness of a disease can be straightforwardly represented by the basic reproduction number. 
However, our findings suggest that the network structure also plays an important role in determining the basic reproduction number, and should be taken into account for assessing the contagiousness.  
In addition, the trend in the effective reproduction number is often simply extrapolated to make short-term predictions about the course of the disease spreading. For example, a rapid decrease in the effective reproduction number towards one is usually interpreted as a sign that the outbreak is approaching its peak and that infections will soon begin to decline. In this work, we have shown that this is not the case in populations with spatial structure, where $R_g$ can stabilise at unity for an extended period before eventually dropping to zero. This suggests that conclusions based on a simple extrapolation may be misguided. We note that the convergence of the effective reproduction number to unity is consistent with the previously observed polynomial epidemic growth in empirical studies~\cite{colgate_risk_1989, szendroi_polynomial_2004, chowell_western_2014, thurner_networkbased_2020}, making spatial structure a plausible explanation for these observations.

The reproduction number gained significant attention during the COVID-19 pandemic. Both the basic reproduction number, $R_0$, and the effective reproduction number, $R_\mathrm{eff}$, have been widely used to evaluate interventions and guide policy decisions, making an in-depth investigation of these measures essential. In our view, $R_0$ has been historically something of an overloaded concept: a single parameter that captures the epidemic threshold, the contagiousness of the disease, its early-stage dynamics, and the final epidemic size. We argue that these different concepts need to be decoupled and carefully examined individually. In particular, $R_0$ is a composite that represents both the contact structure and the intrinsic transmissibility of the disease. Our findings suggest that accurately assessing the stage of an epidemic and anticipating its future trajectory from $R_0$ requires accounting for the underlying contact structure.

\begin{acknowledgments}
J.S. and T.H. are supported by the Strategic Research Council (SRC) established within the Research Council of Finland (Decision Numbers: 364386 and 364371). The calculations presented above were performed using computer resources within the Aalto University School of Science ``Science-IT'' project.
\end{acknowledgments}

\newpage

\section{Appendix}

\end{document}